\documentclass[]{spie}
\usepackage[]{graphicx}
\graphicspath{{figures/}}
\usepackage{verbatim}
\usepackage{amsmath}

\title{Improving the LSST dithering pattern\\and cadence for dark energy studies}

\author{
	Christopher M. Carroll\supit{a,b}, Eric Gawiser\supit{b}, Peter L. Kurczynski\supit{b}, Rachel A. Bailey\supit{b},\\
	Rahul Biswas\supit{c}, David Cinabro\supit{d}, Saurabh W. Jha\supit{b}, R. Lynne Jones\supit{e},\\
	K. Simon Krughoff\supit{e}, Aneesa Sonawalla\supit{f}, W. Michael Wood-Vasey\supit{g},\\
	for the LSST Dark Energy Science Collaboration
	\skiplinehalf
	\supit{a}Dept. of Physics and Astronomy, Dartmouth College,\\6127 Wilder Laboratory, Hanover, NH 03755-3528, USA;\\
	\supit{b}Dept.  of Physics and Astronomy, Rutgers Univ.,\\136 Frelinghuysen Road, Piscataway, NJ 08854-8019, USA;\\
	\supit{c}Argonne National Laboratory,\\9700 S. Cass Avenue, Argonne, IL 60439, USA;\\
	\supit{d}Dept. of Physics and Astronomy, Wayne State Univ.,\\42. W. Warren Ave., Detroit, MI 48202, USA;\\
	\supit{e}Astronomy Dept., Univ. of Washington,\\3910 15th Ave NE, Seattle, WA 98195-1580, USA;\\
	\supit{f}Dept.of Astronomy and Astrophysics, Univ. of Chicago,\\5801 South Ellis Avenue, Chicago, Illinois 60637, USA;\\
	\supit{g}Pittsburgh Particle physics, Astrophysics, and Cosmology Center (PITT PACC),\\Physics and Astronomy Dept., Univ. of Pittsburgh, Pittsburgh PA, 15260, USA
	}

\begin{document}
\maketitle

\begin{abstract}
The Large Synoptic Survey Telescope (LSST) will explore the entire southern sky over 10 years starting in 2022 with unprecedented depth and time sampling in six filters, \emph{ugrizy}. Artificial power on the scale of the 3.5\,deg LSST field-of-view will contaminate measurements of baryonic acoustic oscillations (BAO), which fall at the same angular scale at redshift $z \sim 1$. Using the HEALPix framework, we demonstrate the impact of an ``un-dithered'' survey, in which 17\% of each LSST field-of-view is overlapped by neighboring observations, generating a honeycomb pattern of strongly varying survey depth and significant artificial power on BAO angular scales.  We find that adopting large dithers (i.e., telescope pointing offsets) of amplitude close to the LSST field-of-view radius reduces artificial structure in the galaxy distribution by a factor of $\sim$10. We propose an observing strategy utilizing large dithers within the main survey and minimal dithers for the LSST Deep Drilling Fields. We show that applying various magnitude cutoffs can further increase survey uniformity. We find that a magnitude cut of $r < 27.3$ removes significant spurious power from the angular power spectrum with a minimal reduction in the total number of observed galaxies over the ten-year LSST run. We also determine the effectiveness of the observing strategy for Type Ia SNe and predict that the main survey will contribute $\sim$100,000 Type Ia SNe. We propose a concentrated survey where LSST observes one-third of its main survey area each year, increasing the number of main survey Type Ia SNe by a factor of $\sim$1.5, while still enabling the successful pursuit of other science drivers.
\end{abstract}


%
%
\section{INTRODUCTION}
\label{sec:intro}
The Large Synoptic Survey Telescope (LSST) will observe $\sim$20,000\,deg$^2$ of the sky over 10 years and will make major contributions to astronomy.\cite{SciBook} The main science drivers of LSST include detection of near-Earth objects, the creation of detailed surveys of the Solar System and the Milky Way, enhancing the search for faint transient and variable phenomena, and probing the nature of dark energy and dark matter. LSST will provide four probes of dark matter and dark energy via (1) weak lensing cosmic shear (WL) of galaxies, (2) baryonic acoustic oscillations (BAO) present in the power spectrum of galaxy clustering, (3) the evolution of the mass function of galaxy clusters, and (4) the distance-redshift relationship to Type Ia SNe.

LSST was designed to satisfy the majority of its science goals in one data set (main survey) that will include 90\% of the observing time. The remaining telescope time is then split between observing the LSST Deep Drilling Fields (DDFs) and other Òmini-surveys.Ó The wide range of science objectives leads to competing demands in the design of the main survey. These competing demands can be understood by considering the frequency of visits  (cadence) and the exposure time (depth) dedicated to each part of the sky. In general, studies of time-variable phenomena such as Type Ia SNe prefer high cadence at the expense of uniform depth, while studies that emphasize the detection of spatial correlations, such as BAO in galaxy clustering, prefer uniform depth at the expense of high cadence.

Both cadence and uniformity of depth are impacted by the standard observing technique of taking exposures at small excursions away from nominal survey positions (dithering). A survey design that seeks uniform coverage on the sky inevitably produces areas of overlap between adjacent nominal survey positions. In LSST, these overlap regions are important because they constitute a significant sky-fraction of the survey that has higher cadence than non-overlap regions. By smoothing over areas of unequal coverage, dithering improves uniformity and changes the cadence in these overlap regions.

Our theory of modern cosmology relies heavily upon the cosmological principle, which holds that the universe when viewed from a large enough scale appears homogenous and isotropic from any vantage point. The discovery of an accelerating expanding universe drives us to test this cosmological principle via supernovae studies.\cite{Riess1998} If the universe is indeed isotropic, then the distance-redshift relation traced by light curves of Type Ia SNe observed in different directions on the sky should be identical. However, if the redshift-distance relation varies with position on the sky, the cosmological principle would not hold, and we would be in need of a new or modified theory of cosmic evolution. By using WL and BAO in addition to Type Ia SNe, we can probe for dark energy anisotropy and place tight constraints on the dark energy equation of state.\cite{Ivezic2011}

The intent of our work is to optimize the effectiveness of the LSST observing strategy for dark energy studies while maintaining uniformity in the main survey and limiting any adverse affects on other areas of study. Although the primary focus of our work is Type Ia SNe and survey uniformity, it is worth noting that many of the same considerations taken here also enter when optimizing for studies of weak gravitational lensing and galaxy clusters.\cite{Jee2011,Morrison2012}
%
%
\section{METHODOLOGY AND ANALYSIS GOALS}
\subsection{Operations Simulator}
\label{sec:opsim}
The Operations Simulator (OpSim) was developed in order to verify that LSST can meet all of its required science goals given a specific survey design.\cite{SciBook}  OpSim constructs a detailed ten-year simulated LSST run, providing information on each observation including position on the sky, time and filter, weather conditions, and more. Such detailed models of the LSST ten-year survey are vital for estimating the final co-added depth of the stacked images in each filter as a function of sky position. From this we can determine the overall uniformity of the survey and detect any noticeable patterns or effects that may propagate through our analysis leading to systematic errors. In addition, we can determine the total number and cadence of observations per area of the sky, which is fundamental in determining the efficiency of the LSST observing strategy for dark energy studies as well as the other science drivers. OpSim also accounts for telescope and camera parameters such as slew time, filter exchanges, and scheduled maintenance. These simulations model seeing and weather conditions from on site measurements to predict results as close to actual telescope data as possible.

The LSST original baseline cadence strategy tiles the Southern Hemisphere of the sky with hexagons. Inscribing the hexagon within the roughly circular LSST 3.5\,deg field-of-view (FOV) produces doubly observed areas within an LSST FOV caused by overlaps with neighboring observations. For our analysis we use OpSim run 2.168 which includes 10 Deep Drilling Fields (DDFs). Each DDF consists of a single fixed LSST FOV which is intended for long-term, rapid observations and is planned to be visited more often than the canonical main survey cadence, resulting in increased sensitivity to faint objects. The LSST DDFs will provide rich data sets due to increased time sampling meant for observing extremely faint, distant objects. Our analysis of OpSim excludes these DDFs and focuses specifically on the LSST main survey.

An additional observing strategy is implemented by modifying the OpSim output. This modification is used to post-process the simulation to create large dithers ($\sim$0.5 FOV) that offset the LSST pointing centers on return visits. The dithering pattern that we utilize appears in the output database for OpSim but has not yet been implemented in the Simulator itself. This pattern uses a lattice of points that fills a hexagon inscribed in the LSST FOV.  Each night, the next vertex is chosen as an offset for all telescope pointings versus the center of the hexagon, until all 217 points in the hexagon have been utilized and the pattern begins anew. By the end of the 10 year survey, due to the randomness in which nights a given field is observed, the pattern of dithering vectors applied in each field is close to random and includes shifts as large as the radius of the LSST FOV. These large dithers help to increase survey uniformity by reducing the sharp contrast of overlap regions from their pointing centers. LSST has always planned to utilize some form of dithering to cover the small gaps between CCDs on repeated visits, but the CCD gaps are so small compared to the FOV that this minimal dithering pattern is well approximated by not dithering at all in the current OpSim code, and we will henceforth refer to this as the ``un-dithered" survey. By instead utilizing large dithers we hope to create the most uniform survey possible with negligible negative effects on other science drivers.

\subsection{Analyzing OpSim output with HEALPix}
\label{sec:model}

\begin{figure}[h]
	\begin{center}
		\begin{tabular}{c}
			\includegraphics[height=3cm]{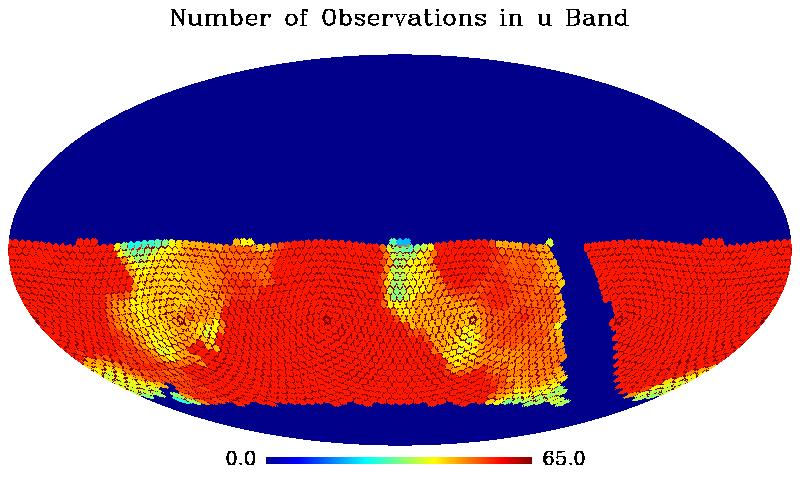}
			\includegraphics[height=3cm]{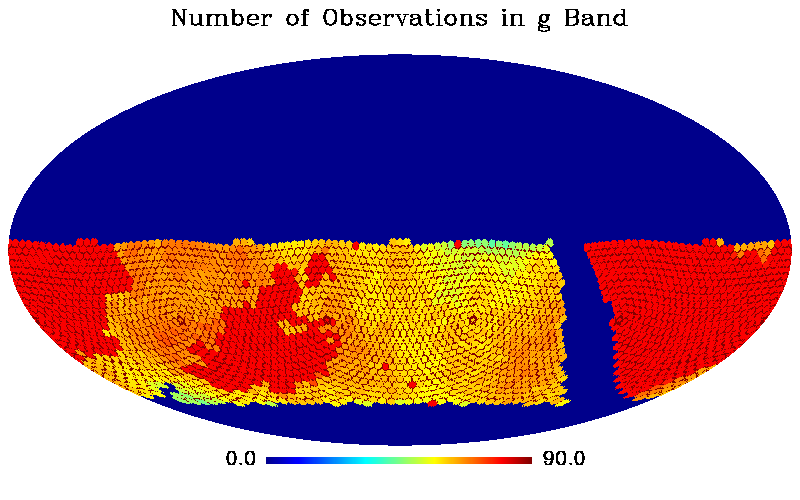}
			\includegraphics[height=3cm]{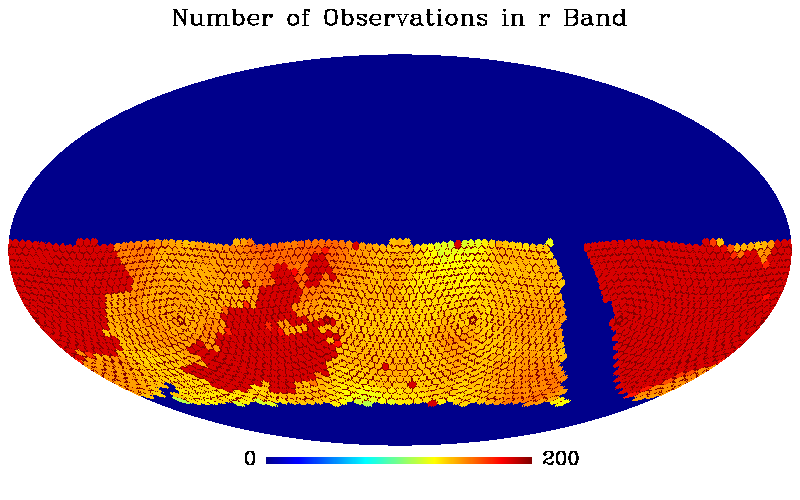}
	   	\end{tabular}
		\begin{tabular}{c}
			\includegraphics[height=3cm]{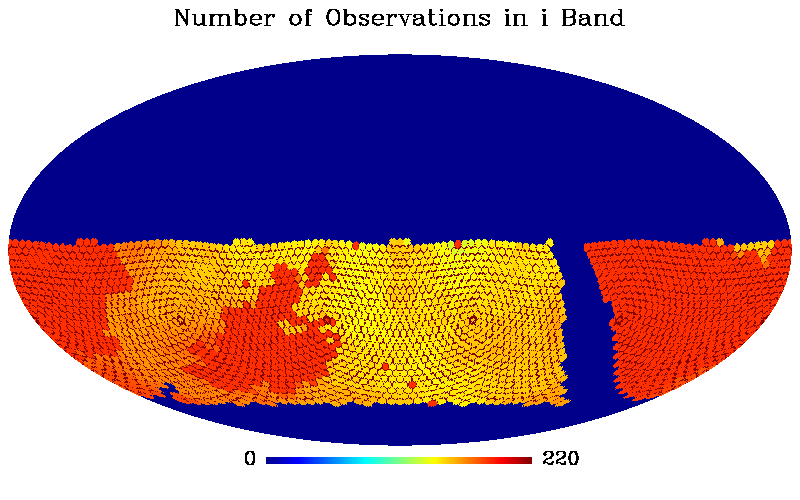}
			\includegraphics[height=3cm]{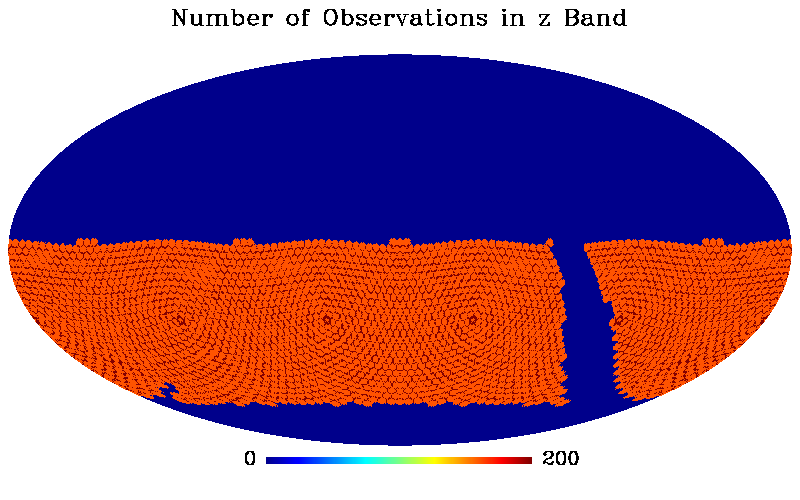}
			\includegraphics[height=3cm]{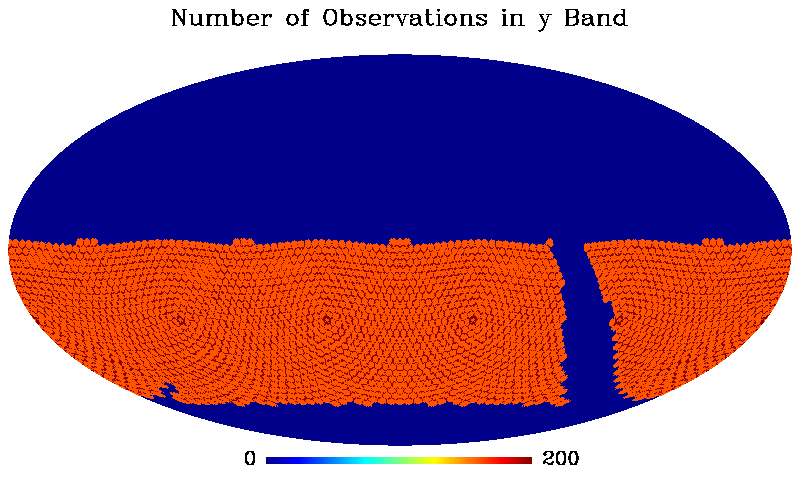}
	   	\end{tabular}
	\end{center}
	\caption[example]
	{\label{fig:num_obs}
Total number of observations in the OpSim un-dithered survey in all bands, \emph{ugrizy}.}
\end{figure}

\begin{figure}[h]
	\begin{center}
		\begin{tabular}{c}
			\includegraphics[height=3cm]{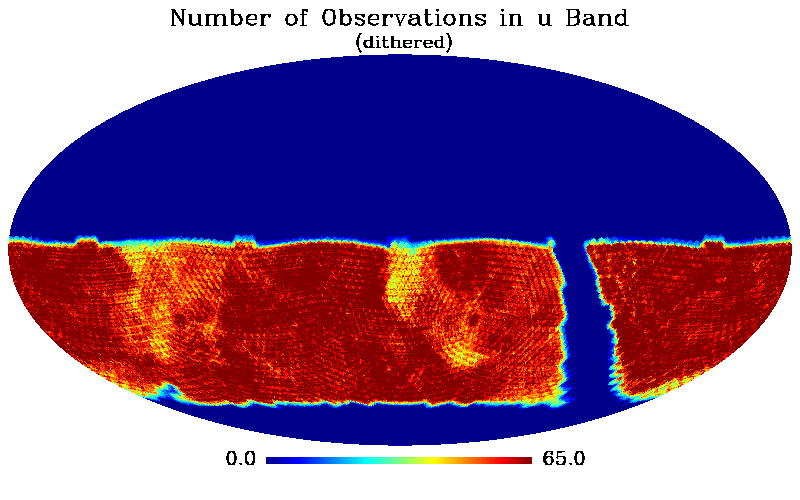}
			\includegraphics[height=3cm]{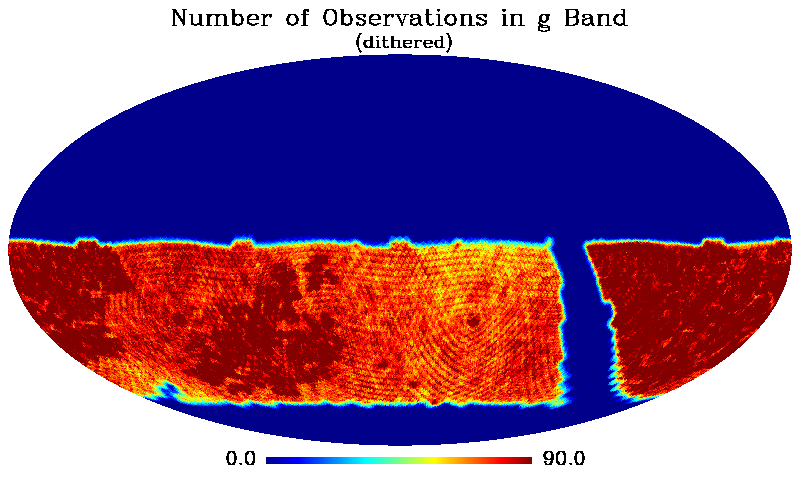}
			\includegraphics[height=3cm]{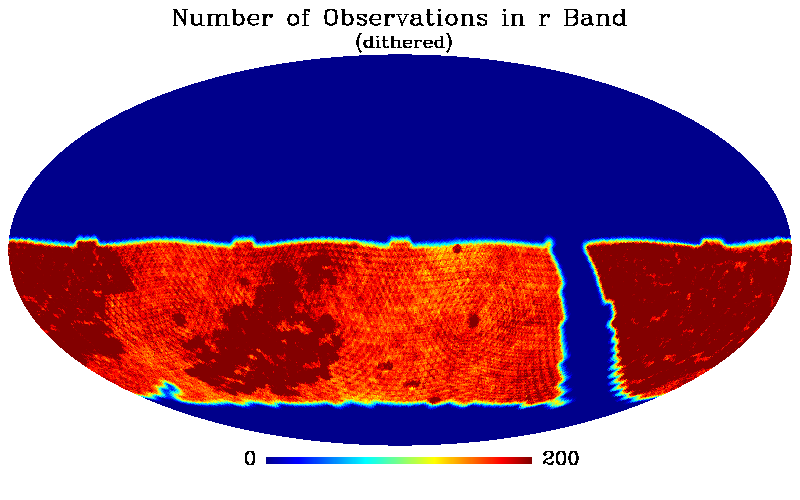}
	   	\end{tabular}
		\begin{tabular}{c}
			\includegraphics[height=3cm]{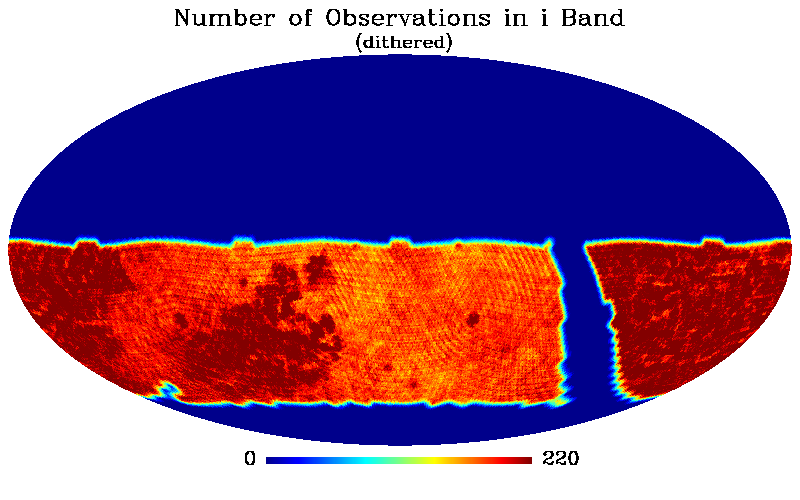}
			\includegraphics[height=3cm]{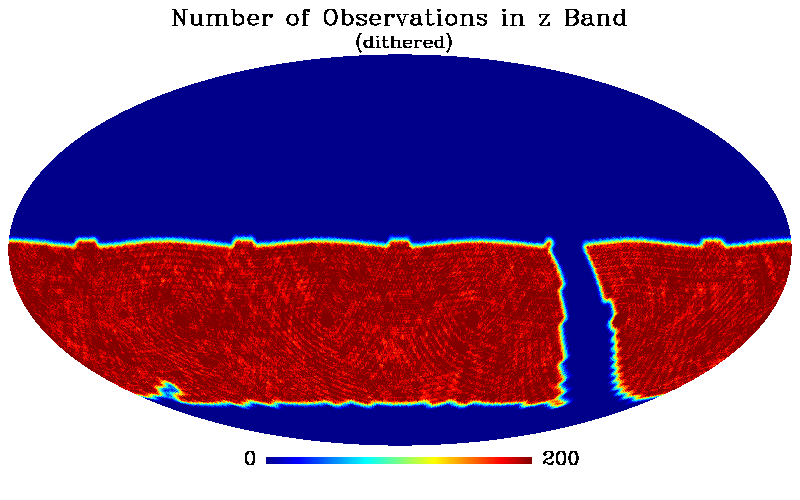}
			\includegraphics[height=3cm]{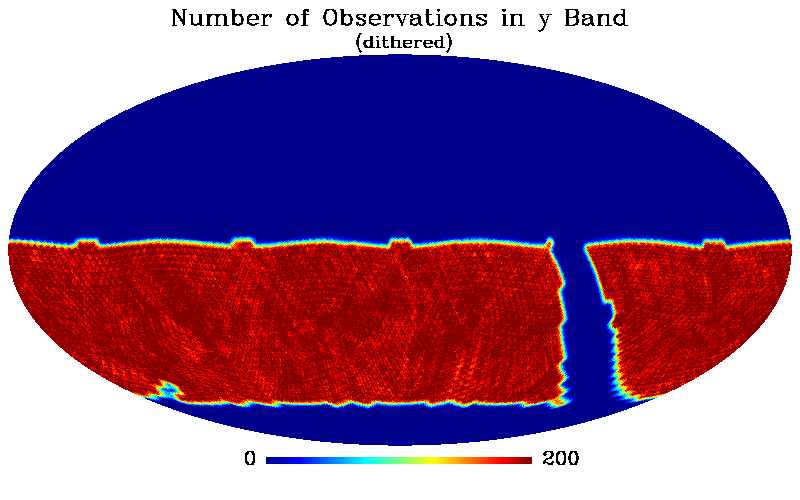}
	   	\end{tabular}
   	\end{center}
   	\caption[example] 
   	{\label{fig:num_obs_dith} 
 Total number of observations in the OpSim dithered survey in all bands, \emph{ugrizy}.}
\end{figure}

The need to model overlap regions prompted us to utilize the Hierarchical Equal Area isoLatitude Pixelization\footnote{http://healpix.jpl.nasa.gov/} (HEALPix) software package\cite{Gorski2005}. HEALPix provides us with a tool for creating uniform tessellation of a sphere with a variable number of equal area pixels. By increasing the number of pixels, and thereby decreasing the area of each pixel, we fill an LSST FOV with an adequate number of pixels to highlight the overlap regions produced by neighboring observations. The total number of pixels is set by the resolution parameter $N_{\mathrm{side}}$ according to the formula $N_{\mathrm{pix}} = 12 \times N_{\mathrm{side}}^2$. Throughout our body of work we have chosen the resolution parameter as $N_{\mathrm{side}}$ = 128, which yields $N_{\mathrm{pix}}$ = 196608, approximately 50 HEALPix pixels per LSST FOV. Each individual HEALPix pixel covers $\sim$0.21\,deg$^2$ and this allows us to make sky maps of OpSim metadata and to observe the overlap regions. Figures \ref{fig:num_obs} and \ref{fig:num_obs_dith} show our findings in all filters \emph{ugrizy} of the main survey and illustrates how large dithers affect the survey, blending the overlap regions and creating a noticeably more uniform survey. Further analysis of the effects of these dithers on survey uniformity will be discussed in Section \ref{sec:results}.

\subsection{Cadence for Type Ia SNe}
\label{sec:cadence}
Most science drivers benefit from an increased number of observations, but to optimize LSST for dark energy studies with Type Ia SNe we also need to increase the cadence of these observations. For Type Ia SNe, light curves require frequent multi-color observations to act as calibrated standard candles, allowing us to probe the farthest edges of the observable universe. We set out to determine the number of HEALPix pixels on the sky with suitable cadence for cosmological studies of Type Ia SNe. Calibrating Type Ia SNe requires an even distribution of observations at $\sim$2-day cadence to capture light curve peaks. A secondary requirement for accurate calibration is observations in multiple bands. With ample filter exchanges, we consider any 60-day period which contains 60 or more observations to be a good observing ÒseasonÓ with adequate cadence for cosmological studies using Type Ia SNe. We seek to determine the total number of good observing seasons in the 10-year main survey. 

%
%
\section{RESULTS}
\label{sec:results}

\subsection{Survey depth and estimated galaxy counts}
\label{sec:gal_count}
Among the numerous outputs provided by OpSim are simulated estimates of the sky brightness and 5-$\sigma$ limiting magnitude for each observation. In our work we use the modified 5-$\sigma$ limiting magnitude output of OpSim, where the 5-$\sigma_{\mathrm{mod}}$ limiting magnitude is calculated by using a V sky brightness model adjusted in each bandpass for moon phase, differences in atmospheric light scatter, and twilight hours to best simulate a real point source detection depth for each visit. We then calculate the stacked 5-$\sigma$ limiting magnitude to estimate the final co-added depth in each HEALPix pixel for that filter; from this we can estimate the number of galaxies we project LSST to find in each pixel. We calculate the co-added depth as
\begin{equation}
	\label{eq:mag_depth}
	\operatorname{5-\sigma}_{\mathrm{stack}} = 1.25\log \sum_{i} 10^{(0.8 \times \operatorname{5-\sigma}_{\mathrm{mod},i})},
\end{equation}
assuming that individual exposures are combined with optimal S/N weighting.\cite{Gawiser2006} Taking the median value in each filter we compare our 5-$\sigma_{\mathrm{stack}}$ values to the LSST science requirements\cite{LSSTReq} for co-added depth per filter in Table \ref{tab:mag_depth}. Using the power-law fit for r-band number counts from MUSYC\cite{Gawiser2006} we determine the estimated number of galaxies to be detected per HEALPix pixel by integrating over all magnitudes brighter than the co-added survey depth in each pixel
\begin{equation}
	\label{eq:numgal}
	N_{\mathrm{gal}} = \int_{-\infty}^{\operatorname{5-\sigma}_{\mathrm{stack}}} 10^{-3.52} \times 10^{0.34m} \,dm.
\end{equation}

\begin{table}[h]
	\caption{Main survey median 5-$\sigma_{\mathrm{stack}}$ depth by filter for the OpSim 2.168 run compared to the LSST Science Requirement Document stretch goal co-added depth. (\emph{left}): Broadband filter used for observation. (\emph{center}): Median depth in the un-dithered and dithered survey respectively. (\emph{right}): LSST SRD idealized co-added depth. It should be noted that any target depths not achieved in OpSim 2.168 are due to over-prioritization of the DDFs within this particular simulation.} 
	\label{tab:mag_depth}
	\begin{center}       
		\begin{tabular}{cccc}
			\hline\hline
			\rule[-1ex]{0pt}{3.5ex}  Filter & Median depth (un-dithered) & Median depth (dithered) & SRD Stretch Goal\\
			\hline
			\rule[-1ex]{0pt}{3.5ex}  \emph{u}  & 26.5 & 26.5 & 26.1\\
			\rule[-1ex]{0pt}{3.5ex}  \emph{g}  & 27.1 & 27.2 & 27.4\\
			\rule[-1ex]{0pt}{3.5ex}  \emph{r}  & 27.3  & 27.4 & 27.5\\
			\rule[-1ex]{0pt}{3.5ex}  \emph{i}  & 26.7 & 26.8 & 26.8\\
			\rule[-1ex]{0pt}{3.5ex}  \emph{z}  & 25.8 & 25.9 & 26.1\\
			\rule[-1ex]{0pt}{3.5ex}  \emph{y}  & 24.6 & 24.7 & 24.9\\
			\hline
		\end{tabular}
	\end{center}
\end{table} 

We modified this integral to correct for incompleteness near the 5-$\sigma$ survey limit as described in Fleming et al. 1995\cite{Fleming1995}. Table \ref{tab:mag_cuts} shows our results on galaxy detection in \emph{r} band. Implementing the proposed large dithers causes a slight increase in the total number of detected galaxies over the course of the survey. There are uncertainties in this process, which translate to uncertainties in the window function estimated in this way. We estimate this window function via the angular power spectrum (Section \ref{sec:power_spec}). Here we assume that the less power in the survey window function, the smaller the uncertainties.

\subsection{Galaxy angular power spectrum}
\label{sec:power_spec}
\begin{figure}[h]
	\begin{center}
		\includegraphics[height=6cm]{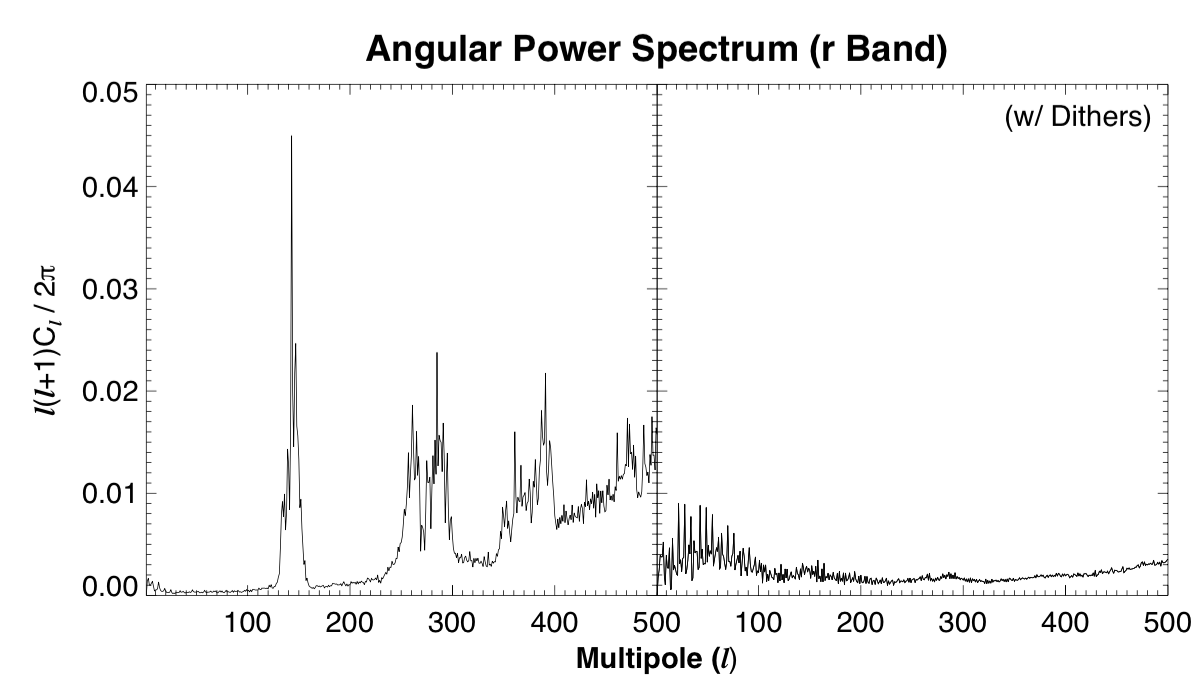}
   	\end{center}
   	\caption[example] 
   	{\label{fig:power_spec} 
Galaxy angular power spectrum (\emph{left}): un-dithered survey, (\emph{right}): with large dithers}
\end{figure}

Characterizing the overall uniformity requires analyzing the spherical harmonic transform of our galaxy detection sky maps. We determine the deviation from survey average ($\Delta N/N_{ave}$) in each pixel and use the HEALPix routine \emph{ianafast} to determine the galaxy angular power spectrum shown in Figure \ref{fig:power_spec}. To ensure a robust, unbiased galaxy sample, we must minimize deviations in the overall co-added survey depth, which creates a more uniform detection limit. From the galaxy angular power spectrum we evaluate the effects that large dithers have on the uniformity of the survey. Comparing the two angular power spectra, it is clear that adding large dithers to the output of OpSim significantly reduces artificial structure in the galaxy distribution. This benefit provided by the implementation of large dithers is crucial for studies of dark energy: note that $\ell \sim 150$ corresponds to an angular diameter of $\sim$$1^\circ$, close to the BAO angular scale at $z \sim 1$. The additional low-$\ell$ power introduced by the large dithers results from the long border of reduced depth that can be seen in Figure \ref{fig:num_obs_dith}. This could be reduced with a careful trimming of that border.
\vspace{.4cm}

\subsection{Magnitude cutoffs}
\label{sec:mag_cuts}
\begin{figure}[h]
	\begin{center}
		\includegraphics[height=7cm]{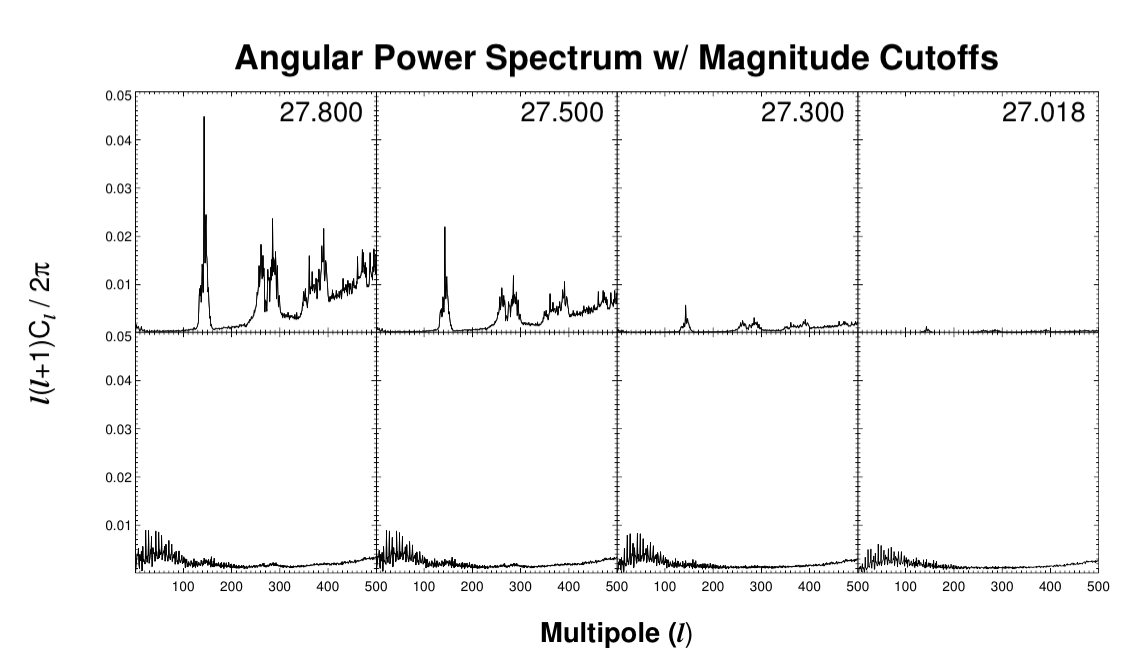}
   	\end{center}
   	\caption[example] 
   	{\label{fig:mag_cuts} 
Galaxy angular power spectra with magnitude cutoffs in (\emph{top}): the un-dithered survey, (\emph{bottom}): with large dithers. The lack of power in the upper right panels results from an oversimplified model for incompleteness. Added low-$\ell$ power in the bottom row comes from reduced depth in survey borders.}
\end{figure}

Further uniformity can be achieved by introducing uniform magnitude cutoffs that smooth variations in the final co-added depth of the heterogeneous survey. Figure \ref{fig:mag_cuts} presents our results for several magnitude cutoffs in the OpSim main survey and their corresponding angular power spectra. Each successive cutoff decreases the total number of galaxies detected in the overall survey, but we find these losses to be mild for cutoffs as bright as $r = 27.3$ (see Table \ref{tab:mag_cuts}). We find a magnitude cutoff of 27.3 to remove sufficient artificial structure in the galaxy angular power spectrum without unnecessary reduction of the LSST galaxy sample size. A more detailed optimization using magnitude cutoffs is the topic of ongoing work.
\begin{table}[h]
	\caption{Total number of detected galaxies after magnitude cutoffs. (\emph{left}): Magnitude cutoff in \emph{r} band. (\emph{center}): Total number of galaxies detected in the un-dithered survey. (\emph{right}): Corresponding number of galaxies detected with large dithers.} 
	\label{tab:mag_cuts}
	\begin{center}       
		\begin{tabular}{ccc}
			\hline\hline
			\rule[-1ex]{0pt}{3.5ex}  Mag. Cut & Num. of Galaxies (un-dithered) & Num. of Galaxies (dithered) \\
			\hline
			\rule[-1ex]{0pt}{3.5ex}  none & 1.41$\times10^{10}$ & 1.46$\times10^{10}$  \\ 
			\rule[-1ex]{0pt}{3.5ex}  27.80 & 1.41$\times10^{10}$ & 1.46$\times10^{10}$ \\
			\rule[-1ex]{0pt}{3.5ex}  27.50 & 1.37$\times10^{10}$ & 1.45$\times10^{10}$ \\
			\rule[-1ex]{0pt}{3.5ex}  27.30 & 1.29$\times10^{10}$ & 1.35$\times10^{10}$ \\
			\rule[-1ex]{0pt}{3.5ex}  27.02 & 1.07$\times10^{10}$ & 1.11$\times10^{10}$ \\
			\hline
		\end{tabular}
	\end{center}
\end{table} 

It is important to note that the most precise probe of dark energy planned for LSST is the joint BAO+WL analysis of bright, $\mathrm{S/N} > 20$, $i < 25.3$ galaxies.\cite{SciBook} Our magnitude cut of $r < 27.3$ is substantially more inclusive than this criterion for typical $r - i =  1$.

\subsection{Cadence results}
\label{sec:cad_results}
\begin{figure}[h]
	\begin{center}
	   	\begin{tabular}{c}
			\includegraphics[height=5cm]{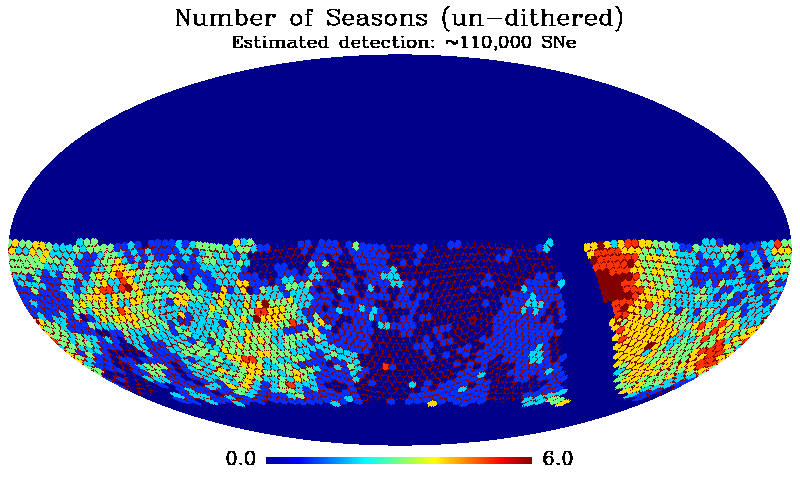}
			\includegraphics[height=5cm]{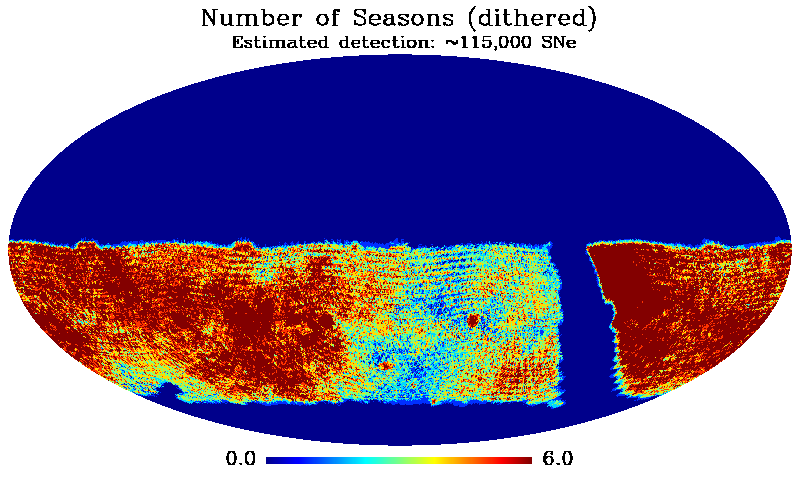}
	   	\end{tabular}
   	\end{center}
   	\caption[example] 
   	{\label{fig:sn_count} 
Total number of good observing seasons in (\emph{left}): the un-dithered survey, (\emph{right}): with large dithers. The total SNe detection counts have been adjusted for edge effects and other cadence issues.}
\end{figure}

For each HEALPix pixel, we determine the number of good observing seasons over the ten-year survey that meet our cadence criteria described in Section \ref{sec:cadence}. Using a conservative limit for supernovae detection rate of 1 SN per square degree per month\cite{Graur2014}, we estimate the total number of supernovae detected over the entire ten-year survey with cadence acceptable for cosmological studies. Our results are presented in Figure \ref{fig:sn_count}. It is important to note that these results are strictly from the main survey and are independent of DDFs, which will contribute an additional $\sim$10,000 high-quality Type Ia SN light curves. We subtract one-third of our SNe detection to account for phase coverage and other cadence issues.

It is also important to note that the main focus of our work is not to determine an absolute number of SNe detected over the length of the survey, rather to show the relative increase in detection via the proposed dithering pattern. We compare the results of OpSim with and without dithers to determine the effects of the overlap regions on supernova count. Large dithers like those proposed for LSST appear to slightly increase the overall number of observations with adequate cadence for cosmological SNe studies. It should be noted that the results presented on supernovae capture are made on the basis of number of visits regardless of filter. Proper calibration of a supernova light curve requires observations in multiple bands. 

The ``cadence output'' can be translated into the input for the supernova analysis package SNANA~\cite{SNANA}.  Here each pixel is used to make an entry in a SNANA simulation library.  The repeated observations of a pixel are time ordered and their observation conditions are taken from the LSST OpSim output to build up a simulation library entry. The new feature here is that this pixelated ``cadence output'' naturally takes into account LSST field overlaps which had previously been ignored in SNANA LSST supernova simulations.  The set of simulation library entries can then be used to used to simulate the observations of various sorts of supernovae (Type Ia and/or core collapse) with various assumptions about dithering to optimize the survey strategy for cosmology with Type Ia SNe.

%
%
\section{DISCUSSION}
\subsection{Time-restricted survey area}
\label{sec:conc_survey}

We propose a drastic change to the scheduled operation of LSST in order to increase cadence and to improve Type Ia SNe capture for dark energy studies. The observational strategy present in OpSim 2.168 covers $\sim$20,000\,deg$^2$ of the Southern Hemisphere of the sky, with a typical gap between visit pairs of 3 nights. Our analysis in Section \ref{sec:cad_results} on the efficiency of LSST to observe Type Ia SNe for cosmological studies has given us hope of a modest population with which to refine our understanding of dark energy. To increase our supernova sample means to increase the cadence of observations of these events\textemdash namely the number of good seasons defined in Section \ref{sec:cadence}. We propose a time-dependent restriction on the observing strategy to concentrate on one-third of the main survey each year over nine years.

Using the analysis methods described in this paper, we determine the cadence of observations for a mock concentrated survey based on OpSim 2.168 where all observations for each year of operation are redistributed by declination. We keep the first year of the default survey untouched and move through each consecutive year, keeping the observations fixed in right ascension and randomly redistribute their declination within the corresponding year's declination bin. Our ``survey" results on the improved cadence and capture of Type Ia SNe using this method are compared to the OpSim run in Table \ref{tab:conc_survey}. We find the concentrated survey produces a factor of $\sim$1.5 increase in Type Ia SNe over the default un-dithered and dithered strategy, and we suspect that the detailed SNANA simulations underway will reveal a larger improvement.
\begin{table}[h]
	\caption{Supernovae detection in the OpSim 2.168 survey and our concentrated survey, along with the improvement factor of SNe detection versus the default un-dithered survey.} 
	\label{tab:conc_survey}
	\begin{center}       
		\begin{tabular}{ccccc}
			\hline\hline
			\rule[-1ex]{0pt}{3.5ex}   & Default survey (un-dithered) & Default survey (dithered) & Concentrated Survey \\
			\hline
			\rule[-1ex]{0pt}{3.5ex} SNe counts & 1.10$\times10^5$ & 1.15$\times10^5$ & 1.70$\times10^5$  \\
			\rule[-1ex]{0pt}{3.5ex} Improvement Factor & 1 & 1.05 & 1.55 &   \\
			\hline
		\end{tabular}
	\end{center}
\end{table}

It should be noted that our coarse approach to a concentrated survey creates artifacts in the survey, such as undesired observations towards the Galactic center. We subtract these false observations from our SNe count for the concentrated survey. Our calculations are therefore missing observations present within the OpSim survey. A full OpSim run would be needed in order to determine more accurate estimates.

\begin{figure}[h]
	\begin{center}
	   	\begin{tabular}{c}
			\includegraphics[height=5cm]{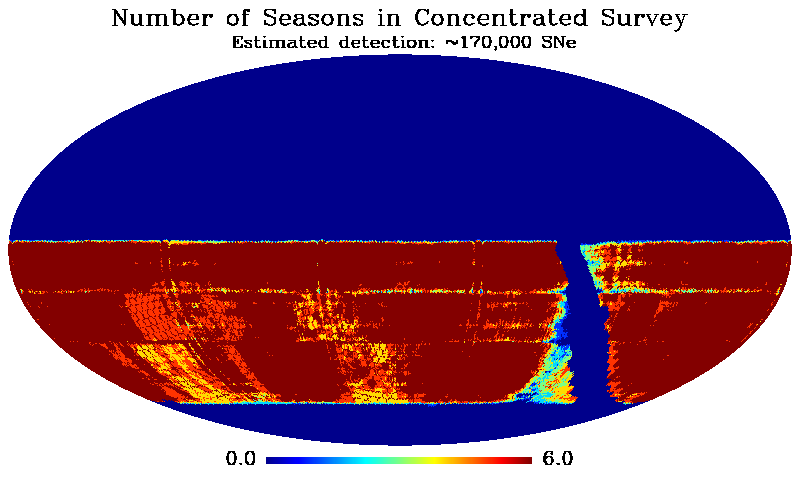}
	   	\end{tabular}
   	\end{center}
   	\caption[example] 
   	{\label{fig:conc_cad} 
Total number of good observing seasons within the concentrated survey. Undesirable observations within the Galactic plane have been removed. The dividing lines in declination between our three equal area regions show up as horizontal lines of reduced cadence.}
\end{figure}

%
%
\section{CONCLUSION}
The astronomical community will benefit from the information LSST collects over its ten-year survey. It is imperative that this survey be optimized to produce maximum results. With a multitude of science drivers, a balance must be struck between competing demands for cadence and survey uniformity. We find each of these survey characteristics to be affected by dithering. Large dithering improves survey uniformity by a factor of $\sim$10 while also modestly improving the cadence over a larger area. Uniform magnitude cutoffs further improve survey uniformity, with a magnitude cut of $r < 27.3$ being roughly optimal to reduce spurious power without a significant loss of statistics. We find that a concentrated survey design, whereby large regions of the sky are emphasized together and uniformity over the entire survey area is achieved at the end of each three years, can significantly improve the yield of Type Ia supernovae. This study illustrates the need for more in-depth analyses of the effects large dithers have on each of the science objectives. Variations on our proposed dithering strategy and the effect on Type Ia SN cadence and survey uniformity is the subject of ongoing work. We caution the reader that our estimates of the yield of Type Ia supernovae are highly approximate and result from an analysis of the total number of LSST observations without regard to which filter those observations were made in. Improved estimates will result from ongoing detailed SNANA simulations of the resulting quality of SN Ia light curves.

%
%
\section*{ACKNOWLEDGMENTS}
We thank Tony Tyson and Beth Willman for their detailed LSST Publications Board review of this article, as well as Pat Burchat and Jeff Newman for helpful comments. This research was supported by grants from the National Science Foundation, AST-1055919, and the Department of Energy, DE-SC0011636.  

%
%


\begin{thebibliography}{10}

\bibitem{SciBook}
{LSST Science Collaborations and the LSST Science Project}, {\em LSST Science
  Book, Version 2.0}, LSST Science Collaboration, Tucson, AZ, 2009.
\newblock arXiv:0912.0201.

\bibitem{Riess1998}
{Riess, A. G. et al.}, ``Observational evidence from supernovae for an
  accelerating universe and a cosmological constant,'' {\em AJ}~{\bf 116},
  pp.~1009--1038, 1998.

\bibitem{Ivezic2011}
{Ivezi\'{c}, \v{Z}. et al.}, ``{LSST: From science drivers to reference design
  and anticipated data products},'' {\em {arXiv.org}} , 2011.
\newblock arXiv:0805.2366v2.

\bibitem{Jee2011}
{Jee, M. J. and Tyson, J. A.}, ``{Towards precision LSST weak-lensing
  measurement - I: Impacts of atmospheric turbulence and optical aberration},''
  {\em PASP}~{\bf 123}, pp.~596--614, 2010.
\newblock arXiv:1011.1913v3.

\bibitem{Morrison2012}
{Morrison, C. B. et al.}, ``{Tomographic magnification of Lyman-break galaxies
  in the Deep Lens Survey},'' {\em MNRAS}~{\bf 426}, pp.~2489--2499, 2012.
\newblock arXiv:1204.2830v3.

\bibitem{Gorski2005}
{G\'{o}rski, K. M. et al.}, ``{HEALPix: A framework for high-resolution
  discretization and fast analysis of data distributed on the sphere},'' {\em
  ApJ}~{\bf 622}(2), p.~759, 2005.

\bibitem{Gawiser2006}
{Gawiser, E. and the MUSYC Collaboration}, ``{The Multiwavelength Survey by
  Yale-Chile (MUSYC): Survey design and deep public ubvriz' images and catalogs
  of extended Hubble Deep Field South},'' {\em ApJS}~{\bf 162}, pp.~1--19,
  2006.

\bibitem{LSSTReq}
{Ivezi\'{c}, \v{Z}. and the LSST Science Collaboration}, ``{Large Synoptic
  Survey Telescope (LSST) science requirements document},'' 2011.

\bibitem{Fleming1995}
{Fleming, D. E. B., Harris, W. E., Pritchet, C. J., Hanes, D. A.}, ``{CCD
  photometry of the globular cluster systems in NGC 4494 and NGC 4565},'' {\em
  AJ}~{\bf 109}, pp.~1044--1054, 1995.

\bibitem{Graur2014}
{Graur, O. et al.}, ``{Type-Ia supernova rates to redshift 2.4 from CLASH: The
  Cluster Lensing And Supernova Survey with Hubble},'' {\em ApJ}~{\bf 783},
  p.~28, 2014.

\bibitem{SNANA}
{Kessler, R. et al.}, ``{SNANA: A public software package for supernova
  analysis},'' {\em {PASP}}~{\bf 121}, pp.~1028--1035, 2009.

\end{thebibliography}
\end{document}